# An extensive Study of Teaching / Learning Quantum Mechanics in College

Bayram Akarsu

Department of Education, Erciyes University, Kayseri, Turkey

Email:bakarsu@erciyes.edu.tr

## Abstract

Quantum physics is considered as one of the most remarkable discoveries of contemporary physics grown during previous century and gradually manifested to the scientific world such as inventions of laser, the transistor, the electron microscope, and semiconductor. Teaching of physical science has been stressed in the *National Science Education Standards (NSES)* from level K-12 as well as many state science standards (Gossard, 2000). The objectives of the current study are to investigate prospective elucidation of the most common learning difficulties, insufficient teaching strategies and other significant instructional or conceptual problems encountered by science and engineering college students at the senior and/or junior level during the instruction of Quantum Physics. Although conceptual issues about learning and teaching of Quantum Physics were addressed in the current study, I mainly focused on the ways the teachers advance while teaching it, as well as considerations of how the classroom environments should function.

**Keywords**: Quantum Physics, Science Education, Teaching and Learning Science.

## I.INTRODUCTION

Everybody studying physics most likely is familiar with the Einstein's impressive eminent saying "God never plays dice" (Ting, 1999) regarding Quantum Physics (QP) or Quantum Mechanics (QM) in its early developments during the World War I. He probably felt this way because he was completely against the fundamental ideas behind Quantum Mechanics and also never convinced in the concept of the "probability concept which constructs the backbone of QP". Of course, Einstein was a believer and believed in God but he was also a determinist like Newton and many other Physicists before him who believe every mechanistic of every dynamics of motion in the universe can be predicted prior to it occurs. Mainly, almost every physicist is strongly convinced in this idea. The most peculiar crash between the development of QP and Einstein's own opinions was that even though Einstein was opposed to the concept of 'probability' he naively contributed to the development of QP with his special and general relativity postulates and their consequences to the rest of the physics issues.

After decades of research and technological advances, Einstein was proven wrong about this new physics, Quantum Theory, by other scientists who didn't share the same ideas about it (Bohr, 1913; Heisenberg, 1925; Schrodinger, 1920). Bohr (1913), Heisenberg (1925), and Schrodinger (1920) developed and applied it to contribute to various new branches of physics such as solid state physics, high energy physics, atomic physics, and molecular physics. When I was first introduced to QP, I was very enthusiastic about it. It proposed enormous and magnificent theoretical ideas no one at that time really believed, such as scientifically correct descriptions of the probability of finding an electron around the nucleus, and how electrons can show both particle and wave properties. These ideas were based on non-deterministic worlds and fundamentally opposite of what Newtonian deterministic physics claims. Many famous physicists including Einstein objected to this new physics because of its counterintuitive suggestions, approaches and explanations of atomic physical phenomena.

The study will first present a brief definition and description of QP. Fundamentally, quantum physics is defined as the physics of the incredibly small and it basically depicts how electrons surround the nucleus of the atom and other subatomic actions. In addition, QP enlightens physicists by successfully elucidating the behaviors of even smaller particles such as electrons, protons, and neutrons. QP has been one of the most important physics born in the 20[th] century because it changed the way physicists examine the nature forever. But its biggest achievement was suggesting indeterminism, probability and non-locality into the foundation of physics. Furthermore, QP is about the characteristics of the subatomic particles and it says that energy is not continuous except in the form of quanta (The term "Quantum" is derived from this word). In modern English, it means "any of the small increments or parcels into which many forms of energy are subdivided' and 'any of the small subdivisions of a quantized physical magnitude (as magnetic moment).
       In the past, scientists accepted as true that light consisted of waves but electrons, neutrons, and protons mostly behave like particles. Einstein discovered that sometimes light possesses particle-like behavior from conclusions of important experiments such as photoelectric effect. Quantum physics applies to the real world in many ways, such as data storage and processors. There were some other important experiments that helped creation of the QP, such as Thomson's experiment with ray tubes that allowed him to discover electrons. Another milestone experiment is Rutherford's with alpha particles and gold which also led to the discovery of the proton. In conclusion, the evolution of QP started with the questions about how light and other particles behave and whether they are wave or particle or both .

After everything settled about QP in the middle of 20[th] century, 50 years after its inception, the physicists focused on teaching the theory of QP to science students rather than its experimental and research-based findings. Furthermore, physics educators proposed many teaching strategies to focus on that issue. For example, Ireson (2000) suggested that teachers in colleges should be sensitive to the variety in the nature of their students' thinking regarding quantum phenomena and that textbook authors and course developers need to draw on the available research to plan a sequence of instruction which allows the student to develop a conceptual framework for a subject that is often counterintuitive to commonsense or mechanistic reasoning. Conversely, whatever they developed they shared very similar teaching techniques in general. The proposed curriculum was mainly based on teacher-oriented classrooms without involving students in the process of learning QP. Basically, even in today's classroom college professors teach QP mostly through direct teaching in a reasonable order by solving a couple of related problems.

The current paper reports on investigation of students' understanding of the concepts of quantum physics. How student reasoning of fundamental concepts and professors' initiatives were probed. Although conceptual issues about learning and teaching of quantum physics were addressed in prior research, the current study primarily explores college professors' opinions and instructional approaches in quantum physics classes, as well as considerations of what the classroom environments are like.

The objectives of the current study were to investigate difficulties of the college students enrolled in introductory undergraduate quantum physics courses with the perspectives of faculty members and students at five big mid-western universities and to explore possible solutions in order to improve understanding of quantum physics for students.

The findings are supported by Kalkanis et al. (1998) and Singh (2001) who proposed solutions for students' insufficient knowledge of mathematical background of quantum physics. The quantum physics curriculum needs to be revised to dedicate four semesters toward this end. Moreover, if the structure of the physics curriculum permits we should study quantum physics concepts over six semesters. Otherwise, students are not able to keep up with the way it is currently handled. To summarize, physics department in colleges should be given the opportunity for spending more time for quantum physics concepts than the one in the current curriculum.

In order to achieve the desired level of students' conceptualization of quantum physics, earlier classes and instructors shouldn't take the whole responsibility. The solution to those problems requires additional courses in the curriculum to prepare students more. This modification can be easily achieved with the aid of two new mathematical physics courses purposefully intended to provide necessary mathematical tools for quantum physics courses. Both courses should be offered to sophomore level students in the physics department and desirably by physics faculty member experts on quantum physics. 16 research articles were reviewed for the current study. Out of 16, seven articles were conducted in pre-university level and remaining nine investigated concepts of quantum physics studied in university classrooms.

In secondary school environment, one of them dealt with teaching strategies for quantum physics course and conceptual difficulties (e.g. abstract side and heavy mathematical content in quantum physics) experienced by the pre-university student in United Kingdom as a teaching model (Ireson, 2000; Muller et al., 2002; Niedderer et al., 1997). Others studied how students make efforts to accommodate the concepts of quantum physics into their conceptual frameworks and the ontological and epistemological status of theoretical entities, and explored students' Implicit or underlying dimensions of reasoning (Mashhadi et al., 1999; Taber, 2003). On the other hand, the second section reviewed literature in college environment. The studies mainly focused undergraduate students' understanding difficulties and misconceptions they experienced in quantum mechanics courses (Ireson, 1999; Singh, 2001).

Present study utilized seven physics professors and over 100 physics students at various colleges in the Mid-Western United States. Out of seven faculty members, five agreed to participate and out of over 200, 86 students returned their questionnaire regarding concepts of quantum physics. We used pseudonyms for faculty members. In addition, course materials and textbooks were examined for the purposes of establishing a standard curriculum.

## II. WHY THIS STUDY IS IMPORTANT

Research studies on students' preconceptions in the area of quantum mechanics, in contrast to other areas of physics, are rarely studied (Singh, 2001). Opposite to classical mechanics, for example, the area of modern physics in high school has little relation to experiences of students in everyday life. Knowledge of and dealing with elements from modern physics do not possess observable phenomenon in students' everyday life.

This current study is an important investigation because it addresses students' perspectives of QP in college level classrooms, which is something that has not been much investigated, which focused on the learning of QP. There were some various important studies that concentrated on learning and teaching key topics of QP in secondary school and college levels such as (Taber, 2004; Singh, 2001; Zollman, Rebello, & Hogg, 2001; Ireson, 1999; Ireson, 2000; Petri and Niedderer, 1998; Johnston, Crawford, & Fletcher, 1998; Roth, 1995).

## III. WHAT THIS STUDY IS NOT

This study does not focus on how well students conceptualize the topics of quantum physics because the intention of this research is to examine what kinds of difficulties students encounter during a QP course. As a researcher, I am allowing the students to feel free to provide what is meaningful to them. In addition, I am giving more attention to students' input about the difficulties they face when learning any quantum physics topics other than investigating whether they understand these topics or not.

Mashaddi et al. (1995) indicated that students usually come across two major problems related to main concepts of quantum physics. First of all, a concept is understood, ultimately, through its relations with other concepts and is the collection of memory elements that are associated with the label (e.g. the photon) and the pattern of their links. Hence, a new concept cannot be explicitly understood until it is linked in a meaningful way to pre-existing concepts (Ausubel, 1963; Novak & Gowin, 1984). The discussion of students' existent conceptions is an important prerequisite for an intended conceptual change (Fischler et al., 1992) and should be included in the current curricula.

Another difficulty regarding teaching for conceptual understanding in QP classrooms as a major goal is evaluating that understanding. Ideally, understanding is a segment of individuals' cognitive structure. However, nobody can guess what is in another's mind so we have to evaluate their performances to gather what cognitive structures they possess. Therefore, investigating conceptual understanding is not an easy task to complete and not a reasonable choice because of its difficulties.

## IV. LITERATURE REVIEW

Several previous studied were examined and inspected. One of the most influential empirical papers which focus on students' understanding and conceptualization of quantum physics belong to Ireson. In one of his paper at Pre-University level, Ireson (2000) suggested a teaching strategy for quantum physics course difficulties experienced by the pre-university student in United Kingdom as a different approach. This study was prominent because it focuses on a different approach to overcoming obstacles, e.g. abstract side and strong mathematical tools in quantum topics, encountered by the students during learning quantum subjects.

Another study by Muller (2002) presented a new research-based course on quantum mechanics in which the conceptual issues of quantum mechanics are taught at an introductory level. This was selected due the fact it focuses on students' misconceptions reported really good information and findings. It had also a huge number of participants (523 high school students in Germany), which make a more significant study. Petri and Niedderer (1998) reported the students' cognitive system for atomic physics as a hypothetical pragmatic model conducted in a German high school with only one participant. This study is the only study investigated by using a qualitative approach about the learning process of an 18-year-old student. Budde, Niedderer, Scott and Leach (2002) conducted a study that utilized "Bremen teaching approach" and made use of this model in order to present an analysis of the learning of two students as they progressed through the teaching unit. This was a part of a big project focusing on the atomic model 'Electronium.' This teaching approach includes the visual Electronium model, as well as the probability model.

Additionally, the studies conducted in secondary level reported similar findings. All of the researchers concluded that learning the concepts of quantum physics is hard because it contains abstract ideas, requires strong mathematical tools, and possesses complicated operations. Students are also experiencing misconceptions such as wave-particle duality and Bohr's model of atom. At the end of his study, Ireson (2000) recommended some ideas to physics instructors: avoid referencing to classical physics, do not introduce photons in the discussion of the photoelectric phenomena, ignore wave-particle duality and statistical interpretation, and finally, avoid introducing the Bohr model of atom when introducing the hydrogen atom.

The second set of the articles was mainly designed to investigate undergraduate students' understanding and difficulties they experienced in quantum mechanics courses. The first study in this section conducted by Singh (2001) sought to analyze the difficulties of advanced undergraduate students in a quantum mechanics course and to compare difficulties and misconceptions. Ireson (1999) conducted the only study I reviewed that used multivariate analysis in his investigation of undergraduate physics students' conceptions of quantum phenomena. Although it is a very small research report, I think it was one of the best research papers published in this area since it manipulate numerous recent papers about modern physics or quantum physics.

Wittmann, Steinberg & Redish (2002) also investigated students' understanding of quantum physics with reporting student reasoning about models of conduction. Although this study dealt with a very specific topic of the quantum physics area, it was well reported and well done with descriptions of the problem. Johnson, Crawford & Fletcher (1998) described student difficulties in learning quantum mechanics. They conducted a study to identify the most important concepts that students need to understand in order to learn quantum mechanics successfully and to recommend the ways the students conceptualize the ideas of quantum mechanics, which makes them difficult. Fischler and Lichtdeldt (1992) conducted another important study that focused on relationships between one of the most important modern physics subjects, the Bohr atomic model, and students' conceptions. The last three studies proposed some instructional models in a quantum theory course. The first one was investigated by Zollman et al (2001) was challenging the abstract difficulty property of quantum mechanics (QM) by creating instructional materials for quantum mechanics. Vokos et al. (2000) have investigated college students' understanding of particle-wave duality in college level physics courses enrolled in quantum physics courses from introductory to advanced levels.

Lei Bao and Edward Redish (2001) conducted a study which focused on understanding probabilistic interpretations of physical systems by two groups of college freshmen and sophomore students. In addition, Cataloglu & Robinett (2001) wrote developed an assessment instrument designed to test conceptual and visualization understanding in quantum theory in order to probe various aspects of student understanding of some of the core ideas of QM. Greson (1999) recommended some useful approaches to the teaching of quantum physics. For example, he suggested the following two approaches: (a) reference to classical physics should be avoided, and (b) teaching of the photoelectric effect should start with electrons, not photons.

Singh (2001) conducted a study investigating the difficulties of advanced undergraduate students toward the end of a full year upper-level quantum mechanics course with concepts related to quantum measurements and time development. Mashhadi and Woolnough (1999) utilized an iterative approach to identify students' conceptions from the data. The types of responses were noted after an initial read-through of the collected responses to a particular question.

In conclusion, every reviewed manuscript mainly focused on undergraduate students' understanding and difficulties they face during quantum physics courses. All of the studies were accomplished with quantitative method. They concluded that students experience difficulties and illustrate deficiencies in quantum physics courses because of the following reasons: (1) insufficient knowledge of particular concepts, (2) heavy mathematical formalism, and (3) the questions related to formulations are not interpreted in the technical practices.

In order to overcome students' difficulties, the researchers suggested some solutions:

- connection with classical mechanics should be avoided
- electrons should be the first topic in the syllabus
- wave-particle duality should be approached

- Heisenberg uncertainty formulism should be introduced at an early stage
- the photoelectric effect should start with electrons, not protons
- the Bohr model of atom should be avoided in the discussion of the hydrogen atom

## V. METHODOLOGY

As noted earlier, this study adopted the questionnaire used in Ireson's article (1999) which focused on investigation of the pre-college level physics students' quantum understanding derived from conceptual statements indicated in his questionnaire. His study addressed high school students' understanding of quantum phenomena. Exploring college students' conceptualization of quantum mechanical concepts and faculty members' approaches to teaching quantum mechanics in their classrooms is the main purpose of the current study, as well as their recommendations to enhance learning its concepts. Besides, it discusses issues related to how I utilized qualitative and quantitative methods with the techniques used.

The purposes of the current study were to investigate difficulties of the college students enrolled in introductory undergraduate Quantum Physics courses with the perspectives of faculty members and students at three big mid-western universities in the US and to explore possible solutions in order to improve understanding of QP for students.

Following specific research questions were raised:

1- What are the difficulties and obstacles that undergraduate students encounter in their QP courses suggested by the faculty members?
2- What are the possible solutions and recommendations to students' difficulties in QP courses by faculty members in the department of Physics?
3- What are the faculty members' beliefs about the course materials (e.g. textbooks, homework, exams, and quizzes etc.) they use during their coursework?

Eighty six undergraduate students and five faculty members in the department of Physics from five big Midwestern universities were selected to serve as subjects in this study. Five undergraduate modern physics, quantum physics, or quantum mechanics classes, and the students enrolled in them, were used for data collection. Both the faculty members and the students who participated in the study volunteered and either teaching or taking one of Quantum Physics, Modern Physics, and Quantum Mechanics classes in the fall and spring semesters of 2006. Gender of faculty members is summarized in Table 1 below.

Table 1.Gender of the Faculty Members and the Students

| Gender | Faculty Member | | Student | |
|---|---|---|---|---|
| | N | % | N | % |
| Male | 7 | 100 | 42 | 49 |
| Female | 0 | 0 | 44 | 51 |
| No Response | 0 | 0 | 0 | 0 |
| Total | 7 | 100 | 86 | 100 |

The ages of participant students ranges from 19 to 21. Most of them are junior or senior year Physics students and the rest are from various science departments such as the Chemistry and Engineering departments (e.g. Computer and Electrical etc.). Each of the students had previously taken at least one mathematical methods course, such as Calculus, differential equations, and complex analysis, etc.

Instrumentation and Data Collection

Faculty members were interviewed at the same time students responded to the questionnaire. Faculty interview protocol questions are attached to the Appendix section. Each interview took about an hour and was audiotaped.

Table 2. Data collection process that took place in this study.

| Data Sources | Research Question | Types of Data Collection Strategies | # of Participant | Data Collection |
|---|---|---|---|---|
| Primary Data Sources | | Interviews Questionnaire | | |
| | Students' conceptual understanding of QP. | Students' Questionnaire | 86 | Students' responses to questionnaire were collected during fall/spring semesters of 2006 |
| | The difficulties students encounter and possible solutions for them. Faculties' beliefs about the course materials. | Faculty Interviews | 5 | All interviews were done during fall/spring semesters of 2006 |
| Secondary Data Supplementary Resources | How class-related materials affect students' learning. | Textbooks, exams, and lab activities | 91 | All supplementary materials were examined thoroughly |

Interviews with five physics faculty members were primary data sources as well as students' responses to the questionnaire. Semi-structured interview questions, adapted from Author (2005) were used. Most of the interview questions focused on students' difficulties of understanding conceptions of quantum physics topics, teaching strategies they use in quantum mechanics classes, and their recommendations in order to increase students' conceptualizations of quantum mechanics. Additionally, regarding students' understanding and conceptualizing about many major quantum mechanics topics and concepts, I utilized a commonly applied questionnaire (Ireson, 1999). The questionnaire mainly included information about many key topics of quantum mechanics and was used to determine if students possess any knowledge of them.

Besides a student questionnaire and faculty interviews, secondary sources were collected to support the purposes of this study. These sources are exams (including midterms, homework, and final exams), textbooks, and laboratory hands-on activities if any. All of the data were collected throughout the academic year of 2005-2006. Interview protocol questions and questionnaires are provided in the appendix section.

## VI. DATA ANALYSIS

The student questionnaire was the main source of data collection that was composed of 29 items to which students will respond on a five-point, strongly disagree to strongly agree, scale. Of 29 questions, 18 determine students' conceptual understanding of quantum phenomena and 11 focus on their conceptual understanding of models. It was adapted from Ireson (1999) because as in previous work with pre-university students (Mashaddi & Woolnough, 1996) and university students (Ireson, 1999), the clustering of students' conceptions were treated as the representative of understanding. This particular questionnaire was selected because the purpose of that study (Ireson, 1999) is similar to the current study. The purpose of quantum physics statements used in the questionnaire was of eliciting students' understanding of quantum phenomena in this study. The students' response to the questionnaire was statistically evaluated by making use of two multivariate techniques, two cluster analysis, and multidimensional scaling, to reveal groups or clusters of responses.

Faculty interview transcripts were analyzed for themes using a constant comparative method and data were reduced into general categories (Glazer & Strauss, 1967). The constant comparative method begins with the researcher searching through data for reoccurring themes or events that can be used as categories to further reduce data. The researcher then attempted to account for the diversity in the data with the developed categories. New categories may be developed or old categories reformulated until a model emerges that describe all the research findings. This process is constant in that it occurs throughout data collection (Bogdan & Biklen, 1998, p.65; Tobin, 2000). Themes were developed from units of data (sentences or paragraphs) that revealed what the difficulties of teaching quantum physics in the classrooms were and what solutions were proposed by the instructors throughout the academic year. Initial themes were formulated from previous research on quantum mechanics professional literature studies (Kalkanis et al, 2002; Petri et al, 1998).

Finally, other data collection pieces consisted of instructors' teaching materials such as textbooks, exam materials, and lab materials in the classroom. The main criteria for analyzing those materials were: (a) are the classroom materials appropriate for grade level? And compared to the most commonly used college quantum physics textbooks? (b) do textbooks include necessary mathematical and background information that students need in quantum physics class? (c) are questions asked in exams appropriate to their level? (d) are students to be given any pop or regular quizzes with early notification? and (e) are necessary formulas and hints provided in exams? The class materials (e.g. textbooks and lab materials) are going to be evaluated based on criteria in 'Guidelines for College Physics Program' report published by the American Association of Physics Teacher (2002).

## VII. FINDINGS

Students' Difficulties of Conceptualizing the Quantum Physics Courses
For the quantitative analysis part of the study, subjects were 86 students who returned the questionnaire and enrolled in either of modern physics, quantum physics or quantum mechanics courses at four big Midwestern and two midsized Eastern universities.

Ireson (1999) investigated undergraduate students' understanding and the results were characterized by the clustering of students' conceptions of quantum mechanics topics. For that reason, Feynman's premise about quantum mechanics 'nothing is deep or accurate' were the central criteria for evaluation of students' conceptual understanding of quantum physical topics. As a result, any findings corresponding to the students' questionnaire were interpreted not at the level of individuals but at the level of the group. Some of the questionnaire statements used for the quantitative part of the study are tabulated below:

Table 3. Some of the statements addressing understanding of quantum phenomena

| | |
|---|---|
| B01 | The structure of the atom is similar to the way planets orbit the sun. |
| B02 | It is possible to have a visual 'image' of an electron. |
| B03 | The energy of an atom can have *any* value. |
| B04 | The atom is stable due to a 'balance' between an attractive electric force and the movement of the electron. |
| B06 | Coulomb's law, electromagnetism, and Newtonian mechanics cannot explain why atoms are stable. |
| B07 | The electron is always a particle. |
| B08 | An atom cannot be visualized. |
| B09 | Light always behaves as a wave. |

Ireson's (1999) study analyzed the students' responses with two multivariate techniques, cluster analysis and multidimensional scaling, to reveal groups or clusters of response and map them onto a Euclidean space symbolizing the structure or dimensions of the responses. Cluster analysis focuses on allocating individuals to a group by utilizing each individual group member, while treating them more like individuals in the same group than those outside the group.

The descriptive statistics table was illustrated based on 86 students and 29 quantum statements in Table 5:

Table 4 .Descriptive analysis of students' responses to the questions

| University | Quantum Thinking (Mean) | Mechanistic Thinking (Mean) | Dual Thinking (Mean) |
|---|---|---|---|
| University A | 3.41 | 2.24 | 3.18 |
| University B | 3.48 | 2.05 | 3.62 |
| University C | 3.25 | 2.47 | 3.80 |
| University D | 3.36 | 2.34 | 3.50 |
| University E | 3.54 | 2 | 3.75 |
| Average | 3.41 | 2.22 | 3.57 |

According to the table 4, students who volunteered the study possess close to the ideal quantum thinking (based on the scale 1 to 5, 1 being disagree and 5 being agree). Therefore, the participants with an average score of 2.1 represents that he/she acquired quantum thinking and this is the ultimate goal to achieve. Similarly, the participants with the average score of 3.5 for quantum thinking symbolized that they didn't achieve desired level of understanding of quantum phenomena but revealed over midpoint (3). For dual understanding, the students showed the desired level of percentages. Students enrolled in individual faculty member's course related to the descriptive results of their ways of thinking are analyzed in the following section.

Cluster two contains statements that favor quantum thinking, for example, 'electrons consists of smeared chare clouds which surround the nucleus', 'orbits of electrons are not exactly determined', and ' whether one labels an electron a 'particle' or 'wave' depends on the particular experiment being carried out.

Cluster three contains statements, fox example, 'electrons are waves' and 'electrons move along wave orbits around the nucleus.' As a result, Table 4 illustrates that 43% of the students demonstrate mechanical thinking with average score of 2.18 out of 5, 70% (average score of 3.48) has quantum thinking, and 68% (average score of 3.38) possessed dual thinking. If students with dual thinking students are assumed to be in the right track, then, more than half of the students 69% participated grasped the ideal thinking in quantum physics class. Hence, faculty members are doing a fine job but in order to increase students' quantum thinking they should revise curriculum, utilize different tools, and maybe spend more time to cover the essential chapters.

Figure 1.Clusters of Statements on Quantum Phenomena

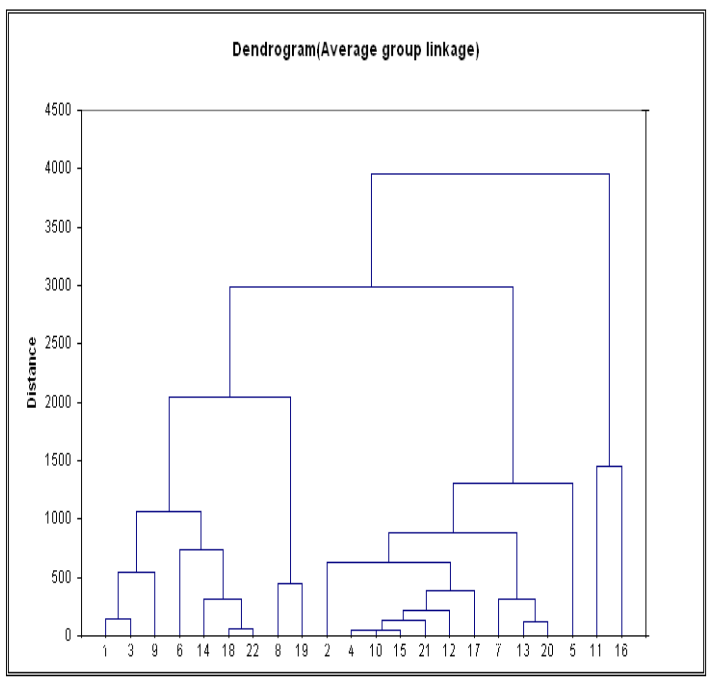

Cluster analysis indicated the groupings of statements or conceptions. The Cluster Analysis using the Complete Linkage method produced the dendogram showing how the statements cluster or group together (see Figure 1)

The clusters were constructed as:

Cluster 1: Mechanistic thinking

B01      The structure of the atom is similar to the way planets orbit the sun.

B02      It is possible to have a visual 'image' of an electron.

B08      An atom cannot be visualized.

B10      In passing through a gap electrons continue to move along straight line paths.

B21      Nobody knows the position accurately of an electron in orbit around the nucleus because it is very small and moves very fast.

B22      It is possible for a single photon to constructively and destructively interfere it self.

B30      If a container has a few gas molecules in it and we know their instantaneous positions and velocities then we can use Newtonian mechanics to predict exactly how they will behave as time goes by.

B25      Electrons move around the nucleus in definite orbits with a high velocity.

B31      During emission of light from atoms electrons follow a definite path as they move from one energy level to another.

B39      The photon is very small, spherical entity.

B35      Electrons are fixed in their shells.

B03      The energy of an atom can have *any* value.

B09      Light always behaves as a wave.

B07      The electron is always a particle.

Cluster 2: Quantum thinking

B04      The atom is stable due to a 'balance' between an attractive electric force and the movement of the electron.

B06      Coulomb's law, electromagnetism, and Newtonian mechanics cannot explain why atoms are stable.

B14      When an electron 'jumps' from a high orbital to a lower orbital, emitting a photon, *the electron is not anywhere in between the two orbits.*

B15      How one thinks of the nature of light depends on the experiment being carried out.

B23     Since electrons are identical it is not possible to distinguish between them.
B27     Electrons move randomly around the nucleus within a certain region or at a certain distance.
B26     When a beam of electrons produces a diffraction pattern it is because the electrons themselves are undergoing constructive and destructive interference.
B36     Orbits of electrons are not exactly determined.
B28     Whether one labels an electron a 'particle' or wave' depends on the particular experiment being carried out.
B33     Individual electrons are fired towards a very narrow slit. On the other side is a photographic plate. What happens is that the electrons strike the plate one by one and gradually build up a diffraction pattern.
B12     The photon is a sort of 'energy particle'.
B18     The photon is a 'lump' of energy that is transferred to or from the electromagnetic field
B16     Electrons move along wave orbits around the nucleus.

Cluster 3 Dual thinking
B13     Electrons are waves.
B19     Electrons consist of smeared charge clouds which surround the nucleus.

## VIII. FACULTY MEMBERS' TEACHİNG STRATEGIES AND THEIR

Albert was teaching a Quantum Mechanics course in the 2006-2007 fall semesters at a big Midwestern university (University A) in the state of Ohio. The course is only offered to senior level students. According to his students' responses to the questionnaire, mean value of classical thinking of them is 2.2 (out of five) and that means approximately 44% of his students do possess mechanical thinking which is almost at the desired level. The ideal score is they shouldn't have any classical thinking because if they do it will be difficult for them to understand the quantum physical topics. In regards with the quantum thinking, they attain a more desired level (68% with quantum thinking).

Albert strongly believed that his students worked really hard regardless of the conceptual barriers of quantum mechanical concepts. He encouraged the students. He also believed that the recitation sessions were really helpful for the students to understand the subject better because teaching assistants helped student with the previous exam questions. In addition, he thought that the pace in the course syllabus is appropriate for the senior level students.

Brian

Brian was another professor teaching a Modern Physics course in the fall semester of 2006 at a big Midwestern university. The course was offered to the junior level students and was a core course for physics students. Questionnaires were handed to all of the students and collected back by the researcher over a two week period. Of all the students in his classroom, only 8 (34%) students responded to the questions in the questionnaire given them. Therefore, his class had the lowest percentage of returned rate among the classes investigated in this study. The interview with Professor B took place in his office in the middle of the fall semester.

His students hold less classical thinking (average 2.05 out of 5.00) than the ones in Albert's classroom and that corresponds to about 40% of the students have the classical thinking. Similarly, the students had an average of 3.48 out of 5.00 of quantum thinking and that is about %69 that is the second highest percentage among all of the faculty members.

Regarding classroom materials, he used the same main textbook as Albert, *Introduction to Quantum Mechanics* (D.J. Griffiths). In addition to this textbook, a software book (Phyla Quantum Physics) which consists of java applets of visual quantum mechanics concepts (such as Wave packets etc.) was recommended as optional supplementary reading. In contrast, he believed that Griffith was the most appropriate textbook for this level of quantum mechanics course. He thought that it had many pros:

"(1) Griffith definitely had an informal way of instruction that made the students engage to the discussions in the book, (2) It is very well written and very clear discussions of the concepts, (3) It consisted of much more exercises compared to the other quantum mechanics textbooks at the end of each section, (4) Finally, it had really good chapter end problems that helped the students to understand how to use the fundamental equations to the applications."

According to Brian, any quantum mechanics or quantum physics class should be studied at a slower pace. He said that he had to cover the first five chapters of the textbook, but he strongly believed that it was way too fast for the students to understand the concepts very well. Concerning students' conceptual understanding of the concepts of quantum physics, he was sharing similar ideas with Albert. Also, he believed that students were having conceptual understanding problems in abstract thinking required for quantum physics such as concepts of measurement and probability of finding particles in an atom.

### Charles

Charles was a professor of physics at a big western university with a BS, an MS, and Ph.D. in physics, so he had a very strong background in the concepts of physics. This first semester quantum mechanics course was taught by another faculty member. His students demonstrated second lowest scores among the other students with 47% of mechanistic thinking and 66% of quantum thinking. It is the lowest percentage of mechanistic thinking and second lowest of the quantum thinking. Akin to Albert and Brian, he was also using *Introduction to Quantum Mechanics* (Griffiths, 2004) as the main textbook material for the class, and unlike them, he didn't have any additional supplementary material.

### David

David was teaching the secondary quantum mechanics class offered to both undergraduate and graduate students in the physics department. During the spring semester of 2006, the students were asked to complete a questionnaire and only four of them returned it. The classical understanding in his classroom has the lowest average compared with the others students (40%) and also the highest percentage of quantum understanding (71%). Overall, his students achieved the highest percentage of desired level of quantum thinking. Like Albert, Brian, and Charles, he was teaching quantum mechanics with the same textbook, Griffith (second edition). Furthermore, the course syllabus mainly included the last chapters in the textbook, which emphasizes Quantum Mechanical applications such as the hydrogen atom, Zeeman Effect, and the EPR paradox/Bell theorem.

David's beliefs with reference to the reasons for students' conceptual understanding of quantum mechanics resemble the previous three faculty members. First, he thinks the students grasp the content of the coursework, but the heavy mathematical tools involved and the essential mathematics is utilized too much and makes students' jobs harder. Second, those mathematical apparatus are mostly novel to the students; therefore, students seem to be taking a math course as well as quantum mechanics.

### Eric

Eric has been teaching different levels of physics courses in colleges for 34 years. The questionnaires were handed to all of the students during the final week of the fall session. His students had the highest average score (3.54) of quantum thinking and lowest score (2.00) of classical thinking. They correspond to 71% and 40%, respectively. Regarding students' conceptual understanding of Quantum Mechanical concepts in Eric's classroom, he claimed that their understanding of the concepts of the quantum mechanics was generally quite good, but their ability to apply this to complex problems varied. Overall, he was satisfied their understanding of the quantum mechanics concepts but worried about the application part. Eric shared similar opinions about students' understanding of the quantum mechanics concepts. He asserted one possible reason for students' understanding difficulties of concepts of the quantum mechanics as: Quantum mechanics is inherently difficult to understand on first exposure because it is counter-intuitive in many ways.

## IX. CLASS MATERIALS

In this study, all of the faculty members interviewed were using the same course material, *Introduction to Quantum Mechanics* (Griffiths, 1998), as the main course textbook material. In addition, some were also using some supplementary materials such as a different textbook, or an interactive book.

Table 5.Course materials

| Faculty Members | Main Textbook Material | Supplementary Material |
|---|---|---|
| Albert | Introduction to Quantum Mechanics (Griffith, D.J.) ($2^{nd}$ edition), Pearson Prentice Hall | A Modern Approach to Quantum Mechanics (Townsend) and v.3 of the Feynman Lectures on Physics |
| Byran | Introduction to Quantum Mechanics (Griffith, D.J.) ($2^{nd}$ edition), Pearson Prentice Hall | Phyla Quantum Physics, An Interactive Introduction (Belloni, M., Wolfgang C., and Cox, A.J.) (with CD-ROM), Pearson Prentice Hall, 2006 |
| Charles | Introduction to Quantum Mechanics (Griffith, D.J.) ($2^{nd}$ edition), Pearson Prentice Hall | None |
| David | Introduction to Quantum Mechanics (Griffith, D.J.) ($2^{nd}$ edition), Pearson Prentice Hall | None |
| Eric | Introduction to Quantum Mechanics (Griffith, D.J.) ($2^{nd}$ edition), Pearson Prentice Hall | None |

Assessment (e.g. midterm, final, quiz, homework) materials the faculty members used were very similar to each other. Their typical assessments consisted of two midterms and one final. Nevertheless, Albert also assigned homework every week. He and Charles, additionally, offered recitation hours to their students for solving and explaining problems in the class and in the exams.

In the faculty members' points of view, classrooms materials are somewhat inadequate but some suggestions to improve them were mentioned: (1) Griffith (2004) has some upsides and downsides; for example it is well written, its content level is superior (e.g. Townsend, 2000) in that engages student with its style, and the examples are suitable for the students (David and Eric) but it lacks instruction, examples, and explanations of the concepts (Albert, Brian, Charles, David), and also it is very peculiar, deep, and hard compared to other QM textbooks (David), (2) Computer simulations and software applications of Quantum Mechanical concepts should extensively be used in order to furnish students' visualization of quantum mechanics (Albert), (3) The syllabus of QM course is usually appropriate to the students levels (Albert) but Brian believes the pace of the syllabus of quantum mechanics is too fast for the students.

## X. CONCLUTION, DISCUSSION AND IMPLICATIONS

In this section, findings related to students' conceptual understanding, proposed solutions by faculty members and classroom materials will be discussed. Then, concluding remarks and implications will be revealed at the end of the chapter. Furthermore, I will provide some suggestions connected to teaching strategies and classroom materials. These suggestions are more than just my ideas about how to improve instruction, because one of the strengths of qualitative data is the richness of the description it provides. Although it was not my goal to collect this type of information, there is evidence present in the data indicating certain strategies are worth trying.

The data analysis illustrated that the hypothesis for the first part of research question one was partially verified by the faculty members. All of the faculty members, except David, supported the idea of abstract nature of quantum mechanics as one of the reasons that makes it more difficult and less understandable by the students. For the second part, only three of them, Albert, Brian, and David, believed that heavy mathematical thinking and tools make quantum physics much harder for the students.

All of the faculty members except David supported the idea of quantum physics having more non-intuitive and abstract concepts than other physics core courses such as Electromagnetic Theory (EM) and Classical Mechanics (CM), and also is a new concept to the students. Albert was agree with David about quantum physics courses require more difficult and abstract mathematical knowledge such as linear and algebra. Brian only supported the first opinion but Charles voted only for the second one. Moreover, Albert was the only professor, who identify quantum mechanics with a bad reputation as a complicated course among college students and so was Charles for quantum physics being very complicated for those senior and junior level students. Among them, only David discussed about some particular concepts that confuse student such as spin and Planck's constant. Finally, Eric shared same opinion with David about come concepts of quantum physics being inherently very hard to comprehend on first exposure to the students

Faculty members' views were used toward answering the second research question, "What are the faculty members' teaching techniques and possible solutions and recommendations by them in the department of Physics?"

Albert proposes more time should be spent on conceptual sections and on clarifying crucial derivations and formulas. Besides, Albert and Brian recommend revision on contents of quantum mechanical courses because, in current schedule, some of the concepts studied in previous modern physics class are repeated again. Through that way, more time can be dedicated to quantum mechanics concepts. He and David firmly stated that physics departments should introduce mathematical concepts to the students before taking any quantum physics courses, such as offering prerequisite mathematical course, and spend more time to solve mathematical calculation problems prior to the midterms. According to Albert and Charles, publishers and authors must develop better curriculum and textbooks for the students. Correspondingly, Brian indicates that quantum mechanics should be studied with a slower pace than in current curriculum so that students could spend more time to grasp conceptual ideas behind it. He, also, suggests that more preliminary courses (modern physics and quantum physics) must be offered preceding quantum mechanics.

For the second research question, my hypothesis was supported partially by the faculty members. Exclusively, they all provided their comments and suggestions about how to design teaching strategies that can improve students' conceptual understanding of concepts of quantum physics. As a final point, we can summarize their opinions about helping students to learn quantum physics better.

For instance, Albert suggested a list of recommendations related to students' success in a quantum physics course: (1) there should be a prerequisite mathematical physics course the students take in order to familiarize themselves with scientific notation and mathematical tools for quantum physics, (2) educators should implement a new version of course curriculum and textbooks for quantum physics courses in order to reduce repetition of the same topics such as Bohr's model of atom studied in modern physics and repeated in quantum physics, (3) the instructor should dedicate more time on delineating conceptual topics of quantum physics and solving mathematical questions, as well as problems asked in examinations to explicate the ideas behind quantum physics theory and to help them to reduce students' difficulties about practicing formulas, (4) last of all, imperative formulas of quantum physics should further be elucidated during class or recitation hours.

The other faculty members mostly shared similar ideas. For example, Charles and David are of the same opinion with Albert about the $1^{st}$ and $2^{nd}$ recommendations. Brian made a different recommendation; quantum physics courses are studied far too quickly currently and that makes students not to comprehend well and to fall behind, so the curriculum should be revised to slower pace with fewer topics.

Ireson (1999) revealed significant consequences and explanations to students' difficulties in quantum physics courses. Figure 1 illustrated two clusters, labeled mechanistic thinking and quantum thinking, generated by cluster analysis, respectively. Cluster one includes statements, for instance, 'the structure of atom is similar to the way planets orbit the sun' and 'nobody knows the position accurately of an electron in orbit around the nucleus because it is very small and moves very fast'.

Discussion

The prior studies reviewed in Chapter 2 supported the findings in this study. I will discuss the finding between the proposed hypothesis and results of the study in the following section.

As discussed in the previous section, the results of the faculty interviews revealed that students mostly struggle in a quantum physics class because of its abstract basics, heavy mathematical formulations, and the various levels of instructions throughout universities in the United States. This was the first research question and its findings are supported by Kalkanis et al. (1998) and Singh (2001). Kalkanis et al. proposed solutions for students' insufficient knowledge of mathematical background of quantum physics. For example, they recommended effective instructional interventions to increase students' knowledge of mathematical background. On the same token, Singh (2001) investigated possible reasons of students' knowledge deficiencies of quantum physics concepts: (1) insufficient knowledge of particular concepts, (2) retrieved knowledge from memory which is not ideally interpreted, (3) knowledge that is retrieved and interpreted at the basic level but cannot be used to draw inferences in specific situations. He concluded that those difficulties cause quantum mechanical misconceptions which were not studied in his article.

In order to improve students' understanding concepts of quantum physics, Zollman et al. (2001) shared similar ideas such as increasingly utilizing interactive computer visualizations and practicing quantum physical problems in recitation hours should be extensively emphasized. He also suggested that hands-on activities and pencil and paper-exercises might boost their knowledge. Johnson, Crawford, and Fletcher (1998) investigated the major difficulties that stand behind college students' deficiencies in a modern physics class. Their finding reflect very similar results with the current study such they also identified quantum physics abstract contents as one of the major causes.

The results of the student questionnaires suggested that approximately 69% of the students acquired an adequate level of quantum thinking which definitely not a desired percentage is. The possible explanation might include strong mathematical tools and operations that prevent them to understand mathematical component of quantum physics. For example, most of the students supported the statement 'if a container has a few gas molecules in it and we know their instantaneous positions and velocities then we can use the Newtonian mechanics to predict exactly how they behave as time goes by' and 'nobody knows the position accurately of an electron in orbit around the nucleus because it is very small and moves very fast'. Both of those statements are false because according to the theory of quantum physics, it is impossible to identify exact position of electrons and the reason for not identifying the position accurately of an electron is because of the Heisenberg uncertainty principle.

On the contrary, almost half of the students supported correct statements, such as, 'electrons move randomly around the nucleus within a certain region or at a certain distance' and 'orbits of electrons are not exactly determined.' As a result, it can be suggested that if instructors do not introduce Bohr's model of atom students won't be confused about the orbits of electrons and do not fall for the incorrect statement. This recommendation is supported by Albert in his suggestion to enhance students' conceptual understanding of quantum physics.

With regard to the course materials the faculty members were using during the study, the Curriculum Guidelines and the Laboratory Guidelines in the *Guidelines for College Physics Program* (American Association of Physics Teachers, 2002) was published to the support of high-quality physics education at the college level. According to curriculum guideline 2 (C-2): 'Instructors should not be limited by the fact that some class time is designed as "lecture" in the timetable. The laboratory component is especially important for any physics course. Well-designed, open ended experiments expose the students to the experimental basis of physics and combine many different skills and concepts.' Also, curriculum guideline 8 suggests that technologies should be implemented in the physics course to help students learn. Additionally, Laboratory guideline (L-1) suggests: 'Laboratory experiences should extend beyond the completion of a recipe of prescribed activities'.

Additionally, curriculum guideline 3 (C-3) offers that 'The objectives of a course should ne clearly articulated, and the course should be assessed regularly by the instructor in the light of students' attainment of the course's objectives.' All of the faculty members prepared a syllabus with clear descriptions of their courses so this guideline was met.

Curriculum guideline 5 (C-5) suggests that 'The mathematical and conceptual level of any physics course must be consistent with the abilities of the students in that course.' In addition, curriculum guideline 13 (C-13) implies that different courses entirely in the same class time with the same instructor should be avoided. Those two guidelines are related to the students' mathematical dilemma in quantum physics courses. All of the faculty members, except Professor E, complained about the complex mathematical tools required for quantum physics but only Albert and David suggested to offer additional mathematical physics course prior to quantum physics course.

## XI. IMPLICATIONS FOR INSTRUCTIONS

A series of implications arise from the data presented in findings and result sections which are examined in the following section. First, faculty members complained about the conceptual problems of students and categorized them into mathematical and abstract physical difficulties. All of them suggested three recommendations regarding mathematical complexities of quantum physics as follows: Students do not connect formal mathematical training and thinking such as algebra and calculus with necessary mathematical tools for quantum physics such as complex algebra and partial differential equations; Students struggle with the new mathematical instrument and notation used in quantum physics; Students have problems with the mathematical formulations of quantum physics.

For abstract side of quantum physics, as reported by Singh (2001) and Wittmann et al. (2002) concerning students' difficulties, quantum physics consists of more non-intuitive and abstract concepts than other physics topics and that confuses students. This result was confirmed by our study. Our study also showed that students get confused with the hard concepts of quantum physics which is considered to be inherited and as suggested by Ireson (1999) and Greson (1999), concepts linked to classical physics should be avoided. Besides, two important topic of difficulty for students from new concepts of quantum physics were spins and Planck's constants.

The list of troubles students are experiencing related to the abstractness of quantum physics produces numerous concerns that need to be elucidated in the context of instruction. The mathematical concern is the most important problem that needs to be addressed. In order to prevail over this difficulty, as Kalkanis et al. (2002) suggested, faculty members should link students' prior mathematical concepts with the one necessary for quantum physics, or if students do not possess the required mathematical knowledge then those concepts should be introduced at the beginning of the course. Alternatively, offering a mathematical course specifically designed for physics students who are going to take quantum physics could solve this problem, too.

Sadly, the data collected in this study revealed that colleges are not offering enough courses to prepare physics students for quantum courses. Only two of five universities investigated in this study offer such a course like "Mathematical Physics" every semester for junior level physics students. I strongly believe that although it is good to offer such courses, it is too late because they take a quantum physics course in the same year and only have one more year to graduate. Therefore, at least one mathematical course intended for quantum physics needs to be offered for students in their sophomore year so they would have one year to digest them and be ready for the heavy and strong mathematical tools and operations in quantum physics courses in the following semester.

The second critical dilemma is the abstract ideas behind quantum physics and its connection among other physics theories. My major findings were aligned with two previous studies that I reviewed (Ambrose et al, 1999; Bao & Redish, 2002). It was concluded by Bao and Redish (2002) that students experienced difficulties in quantum mechanics courses because of their weak background in classical mechanics. Ambrose et al. (1999) discovered that student in a modern physics course articulated ideas about the wave-particle duality of light. In conclusion, students' deficiencies of understanding the concepts of classical physics persist as they progressed though curriculum and generated difficulties in the more advanced courses like quantum physics.

In order to achieve the desired level of students' conceptualization of quantum physics, earlier classes and instructors shouldn't take the whole responsibility. The solution to those problems requires additional courses in the curriculum to prepare students more. This modification can be easily achieved with the aid of two new mathematical physics courses purposefully intended to provide necessary mathematical tools for quantum physics courses. Both courses should be offered to sophomore level students in the physics department and desirably by physics faculty member experts on quantum physics. Also, they should include sections that connect pure mathematics to math applications in quantum physics. Some textbooks, like Ross (1984), can help students to prepare them for essential mathematical background of quantum physics.

As Brian stated, the pace in quantum physics courses is beyond the level students comprehend. One can debate whether quantum physics courses (Modern Physics, Quantum Physics, and Quantum Mechanics) should be offered to students in a more extended period of time. Overall, participated faculty members strongly support the idea of devoting more time to instruct the concepts of quantum physics courses so curriculum need to revised to dedicate four semesters toward this end. Moreover, if the structure of the physics curriculum permits we should study quantum physics concepts over six semesters. Otherwise, students are not able to keep up with the way it is currently handled. Once they are behind in the curriculum, there is no way they can catch up because they need to learn adequate fundamental chapters in order to understand more advanced chapters. To summarize, physics department in colleges should be given the opportunity for spending more time for quantum physics concepts than the one in the current curriculum.

List of References

[1] Author (2004, April). *Quantum mechanics: Is it really hard or...?*. Paper presented at the annual meeting of the National Association of Research in Science Teaching, Dallas.

[1] Ambrose, B., Shafer, P., Steinberg, R., & McDermott, L. (1999). An investigation of student understanding of single-slit diffraction and double-slit interference. *American Journal of Physics, 67* (2), 146-155.

[1] American association of physics teachers. (2002). *Guidelines for College Physics Program*. College Park, MD.

[1] Axmann, W.J. (2002). Research, *development, and preliminary testing of interactive engagements for teaching quantum mechanics to undergraduate physics majors*. Unpublished doctoral dissertation, University of Kansas.

[1] Ausubel, D. P. (2000). *The acquisition and retention of knowledge*: *A cognitive view*. Dordretch: Kluwer.

[1] Bao, L., & Redish, E.F. (2001). Understanding probabilistic interpretations of physics systems: a prerequisite to learning quantum physics. *American Journal of Physics, 70* (3), 201-217.

[1] Bao, L., & Redish, E. (2002). Understanding probabilistic interpretations of physical systems: A prerequisite to learning physics. *American Journal of Physics, 70*(3), 210-217.

[1] Belloni, M., Christian, W., & Cox, A.J. (2006). *Physlet Quantum Physics, an Interactive Introduction* (with CD-ROM). Pearson Prentice Hall.

[1] Bogdan, R. C., & Biklen, S. K. (1998). *Qualitative Research for Education* (3rd Ed.). Boston: Allyn and Bacon.

[1] Bohr, N. (1913). On the Constitution of atoms and molecules. *Philosophical Magazine, 26* (6), 1-25.

[1] Budde, M., Niedderer, H., Scott, P., & Leach, J. (2002). The quantum atomic model 'Electronium': a successful teaching tool. *Physics Education, 37* (3), 204-210.

[1] Cataloglu, E., & Robinett, R.W. (2002). Testing the development of student conceptual and visualization understanding in quantum mechanics through the undergraduate career. *American Journal of Physics, 70* (3), 238-251.

[2] Chiang, S. H. (1986). *Individualized instruction for junior high school physics in southern Taiwan*. Unpublished doctoral dissertation, Indiana University, Indiana.

[3] Eberly, J. (2002). Bell inequalities and quantum mechanics. *American Journal of Physics, 70* (3), 276-279.

[4] Feynman, R. (1985). *Quantum Electrodynamics: the strange theory of light and matter*. London: Penguin.

[5] Feynman, R. (1985). *Students' conceptions and constructivist teaching approaches. Improving Science Education* (pp. 46-49). Chicago: The University of Chicago press.

[6] Feynman, R. (2004). *Feynman Lectures on Physics Volumes 3*. Basic Books; Unabridged edition.

[7] Fischler, H., & Lichtfeldt, M. (1992). Modern physics and students' conceptions. *International journal of science education, 14* (2), 181-190.

[8] Flecther, P., & Johnston, I. (1999, March). *Quantum mechanics: Exploring conceptual change*. Paper presented at the annual meeting of the National Association of Research in Science Teaching, Boston.

[9] Frederick, C. (1978). A mechanical model for teaching quantum mechanics. *American Journal of Physics, 46* (3), 242-247.

[10] Gardner, D. E. (2002). *Learning in quantum mechanics*. Unpublished doctoral dissertation, Purdue University, Indiana.

[11] Gisin, N., & Rigo, N. (1995). Relevant and irrelevant nonlinear Schrödinger equations. *Journal of Physics A: Mathematics General*, 28, 7375-7390.

[11] Gossard, (2000). Students' conceptions in quantum physics. *American Journal of Physics, 74* (3). 345-357.

[12] Griffiths, D.J. (2004). *Introduction to Quantum Mechanics* (2nd edition). Benjamin Cummings.

[13] Heisenberg, V. W. (1925). Over the relationships between quantum theory and classical mechanics. *Time writing for Physics, 33*, 879-893.

[14] Herrmann, F. (2000). The Karlsruhe Physics course. *European Journal of Physics, 21*, 49-58.

[15] Ireson, G. (2000). The quantum understanding of pre-university physics students. *Physics Education, 35*(1), 15-21.

[16] Ireson, G. (1999). A multivariate analysis of undergraduate physics students' conceptions of quantum phenomena. *European Journal of Physics*, 20, 193-199.

[17] Ireson, G. (1999). The quantum understanding of pre-university physics students. *Physics Education*, *35* (1), 15-21.

[18] Johnson, I.D., Crawford, K., & Fletcher, P. R. (1998). Student difficulties in learning quantum mechanics. *International Journal of Science Education*, *20* (4), 427-446.

[19] Johnson, S., & Gutierrez, T. (2002). Visualizing the phonon wave function. *American Journal of Physics, 70* (3), 227-237.

[20] Kalkanis, G., Hadzidaki, P., & Stavrou, D. (2002). An instructional model for a radical change towards quantum mechanics concepts. *Science Education*, *90* (3), 257-280.

[21] Kroemer, A. (1994). Investigating quantum physics in the high schools environments. *Physics Education, 56* (3), 127-139.

[22] Lawrence, I. (1996). Quantum physics in school. *Physics Education, 73* (7), 278-286.

[23] Martin, J. L. (1974). Quantum mechanics from the cradle? *Physics Education, 9* (7), 448-451.

[24] Mashhadi, A. (1996). *Students' conceptions of quantum physics.* Research in Science Education in Europe. London: Falmer.

[25] Mashhadi, A., & Woolnough, B. (1997, June). *Dualistic thinking underlying students' understanding of quantum physics.* Paper presented at the annual meeting of the International Conference on Thinking, Singapore.

[26] Mashhadi A., & Woolnough B. (1999). Insight into students' understanding of quantum physics: visualizing quantum entities. *European Journal of Physics, 20,* 511-516.

[27] Mermin, D. N. (1985). Is the moon there when nobody looks? Reality and the quantum theory. *Physics Today, 38* (4), 38-47.

[28] Muller, R., & Wiesner, H. (2001). Teaching quantum mechanics on an introductory level. *American Journal of Physics, 70* (3), 200-209.

[29] Novak, J. D., & Gowin, D.B. (1984). *Learning how to learn.* Cambridge, England: Cambridge University Press.

Olsen, R.V. (2002). Introducing quantum mechanics in the upper secondary school: a study in Norway. *International Journal of Science Education, 24* (6), 565-574.

[30] Petri, P. & Niedderer, H. (1998). A learning pathway in high-school level quantum atomic physics. *International Journal of Science Education, 20* (9), 329-347.

[31] Roth, W.M. (1995). Affordances of computers in teacher-student interactions: the case of Interactive Physics. *Journal of Research in Science Teaching, 32,* 329-347.

[32] Schrodinger, E. (1920). Outline of a theory of color measurement for daylight vision. *Physics Annual, 63* (4), 397-520.

[33] Singh, C. (2001). Student' understanding of quantum mechanics. *American Journal of Physics, 69* (8), 885-895.

[34] Steinberg, R. N., Oberem, G. E., & McDermott, L. C. (1996). Development of a computer-based tutorial on the photoelectric effect. *American Journal of Physics, 64*(11), 1370-1379.

[35] Stuewer, R. H. (2000). The Compton effect: transition to quantum mechanics. *Annual Physics, 9,* 975-989.

[36] Taber, K.S. (2002). Conceptualizing quanta: illuminating the ground state of student understanding of atomic orbital. *Chemistry Education: Research and Practice in Europe, 3* (2), 145-158.

[37] Taber, K.S. (2002). Conceptualizing quanta: illuminating the ground state of student understanding of atomic orbital. *Chemistry Education: Research and Practice in Europe, 3* (2), 159-173.

[38] Taber, K.S. (2004). Learning Quanta: barriers to stimulating transitions in student understanding of orbital ideas. *Science Education,* 94-116.

[39] Ting, S.C.C. (1999). *The bible according to Einstein: a scientific complement to the holy bible for the third millennium* (1st Ed.). Jupiter Scientific Publishing Company.

[40] Tobin, K. (2000). Research on instructional strategies for teaching science. In D. Gabel (Ed.), *Handbook of research in science teaching and learning.* New York: Macmillan.

[41] Townsend, J.S. (2000). *A Modern approach to quantum mechanics.* University Science Books; 2Rev Edition.

[42] Vokos. S., Shaffer, P.S., Ambrose, B.S., & McDemott, L.C. (2000). Student understanding of the wave nature of matter: Diffraction and interference of particles. *Physics Education, 68* (7), 42-51.

[43] Wittmann, M.C., Steinberg, R.N., & Redish, E.F. (2001). Investigating student understanding of quantum physics: Spontaneous models of conductivity. *American Journal of Physics, 70* (3), 218-226.

[44] Zollman, D.A., and Rebello, N.S., & Hogg, K. (2001). Quantum mechanics for everyone: Hands-on activities integrated with technology. *American Journal of Physics, 70*(3), 252-260.